# Comments on Extraction of Work from a Single Thermal Bath in the Quantum Regime

Elias P. Gyftopoulos
Massachusetts Institute of Technology
77 Massachusetts Avenue, Room 24-111
Cambridge, Massachusetts  02139 USA
and
Michael R. von Spakovsky
Virginia Polytechnic Institute and State University
Blacksburg, Virginia  24061 USA

In a PRL [1], the authors claim to show that "the Clausius inequality can be violated, and that it is even possible to extract work from a thermal bath by cyclic variations of a parameter ("perpetuum mobile"), and that the physical cause for this behavior is traced back to quantum coherence in the presence of the nonequilibrium bath (sic)".

Our evaluation of the letter leads us to the unequivocal conclusions that the claims are false because they lack logical consistency, use either undefined or erroneous terminology, and misrepresent the foundations of thermodynamics.  For example, in the inequality đ$Q \leq T$d$S$, đ$Q$ is the energy of a heat interaction, but it is not clear whether $T$ and d$S$ refer to the reservoir with which the interaction occurs, to the system, or one to the system and the other to the reservoir.

Without reference to quantum theory, a logically consistent, unambiguous, noncircular, and complete exposition of thermodynamics [2] reveals that any system (both large and small, including a one spin system), in any state (both thermodynamic equilibrium and not thermodynamic equilibrium) has a nonstatistical instantaneous property called entropy in the same sense that any state has inertial mass as a nonstatistical instantaneous property.

Among the many theorems that entropy must satisfy, one is the impossibility of a perpetual motion machine of the second kind (PMM2).  The reason is that a PMM2 is required to extract only energy from a system under conditions that demand the extraction of both energy and nondestructible entropy [3].

The same conclusion is reached by using a unified nonstatistical quantum theory of mechanics and thermodynamics [4, 5] which reveals that quantum coherence is not relevant to the impossibility of a PMM2.  Of course, if a system is in a state that is not thermodynamic equilibrium, it is possible to extract only energy because then there exist states with lower energy and the same or larger entropy than the initial state.  However, under such conditions the system is not and cannot be characterized as a bath (reservoir), and neither any laws nor any theorems of thermodynamics are violated [3].  The precise definition of a reservoir is given in reference [6].